\documentclass[10pt,letterpaper]{article}
\usepackage[top=0.85in,left=2.75in,footskip=0.75in]{geometry}

\usepackage{amsmath,amssymb}

\usepackage[table]{xcolor}

\usepackage{array}

\usepackage{authblk}
\usepackage{lastpage,fancyhdr,graphicx}

\begin{document}


\date{}

\title{Network Analysis of Global Banking Systems and Detection of Suspicious Transactions}

\author[1]{Anthony Bonato}
\author[2]{Juan Chavez Palan}
\author[1]{Adam Szava}

\affil[1]{Department of Mathematics, Toronto Metropolitan University, Toronto, Ontario, Canada}
\affil[2]{CIBC Capital Markets, Toronto, Ontario, Canada}
\maketitle

\begin{abstract}

A novel network-based approach is introduced to analyze banking systems, focusing on two main themes: identifying influential nodes within global banking networks using Bank for International Settlements data and developing an algorithm to detect suspicious transactions for anti-money laundering. Leveraging the concept of adversarial networks, we examine Bank for International Settlements data to characterize low-key leaders and highly-exposed nodes in the context of financial contagion among countries. Low-key leaders are nodes with significant influence despite lower centrality, while highly-exposed nodes represent those most vulnerable to defaults. Separately, using anonymized transaction data from Rabobank, we design an anti-money laundering algorithm based on network partitioning via the Louvain method and cycle detection, identifying unreported transaction patterns indicative of potential money laundering. The findings provide insights into system-wide vulnerabilities and propose tools to address challenges in financial stability and regulatory compliance.
\end{abstract}

\section*{Introduction}\label{intro}

Networks play a central role in understanding complex systems, ranging from social interactions to financial systems. In the context of global banking, network analysis offers powerful tools to uncover hidden patterns, assess systemic risks, and develop strategies for regulatory compliance. The interconnected nature of financial institutions means that vulnerabilities in one part of the system can propagate, leading to cascading effects. These dynamics are particularly relevant in two key areas: identifying influential players within banking networks and detecting potentially illicit activities such as money laundering.

The Bank for International Settlements (or BIS) provides consolidated data on cross-border banking relationships, offering a unique opportunity to study the global banking network. This data has been used to model risk contagion and identify critical nodes whose influence extends far beyond their direct connections. Previous studies leveraging BIS data have demonstrated its crucial role in modeling cross-border banking exposures, where network-based methods were employed to analyze systemic risk and potential contagion in global financial systems \cite{cerqueti,chen1,chen2,hughes, marcussen}. In addition, centrality measures such as the Common Out-neighbor (or CON) score and PageRank have been used to detect alliances, leaders, and other influential nodes in adversarial or competition-based network models, shedding light on the dynamic interplay of financial institutions \cite{bonato6,bonato2,bonato3}. 

In this study, we analyze the BIS data using the framework of adversarial networks, which represent competitive and antagonistic relationships between nodes. We leverage centrality measures such as the CON score and PageRank to uncover two types of influential players: low-key leaders, whose influence is understated by traditional centrality measures, and highly-exposed nodes, which are particularly vulnerable to financial contagion.

In addition to systemic risk, the global banking network is also a critical battleground for combating money laundering. Anti-money laundering (or AML) efforts often rely on transaction monitoring systems that flag suspicious activity based on predefined thresholds or supervised machine learning models; see \cite{cassella,nojeim,schneider,white}. However, these approaches face significant challenges, including the scarcity of labeled data and the evolving tactics of money launderers. While supervised learning techniques require accurate historical data on suspicious transactions, such data is often limited due to privacy concerns or regulatory restrictions. This scarcity creates an inherent bias in machine learning models, which are unable to detect novel money laundering patterns that deviate from known typologies. Furthermore, reliance on predefined thresholds can be exploited by money launderers through techniques such as structuring, where transactions are divided into smaller amounts to avoid detection. These challenges have prompted the recent exploration of approaches such as community detection and cycle detection to identify suspicious transaction patterns, particularly those involving sub-threshold amounts or complex layering tactics; see \cite{bonAML,italy}.

In an effort to enhance AML efforts, we propose a network-based algorithm for detecting unusual transaction patterns using anonymized data from Rabobank. Our approach integrates community detection via the Louvain algorithm and cycle detection, enabling the identification of potential money laundering schemes without prior knowledge of flagged accounts. By focusing on unreported transactions under regulatory thresholds and leveraging network structure to detect circular transaction patterns, our method aims to enhance the ability of financial institutions to uncover hidden layers of suspicious activity. Combining community detection with cycle identification ensures scalability to large transaction networks, making it a practical tool for the initial screening of high-risk accounts.

Our work bridges the study of global banking networks and AML techniques, highlighting the value of network analysis in addressing two critical challenges. It provides insights into systemic vulnerabilities by identifying key players who influence banking stability and proposes practical tools to enhance regulatory efforts in detecting financial crime. By combining theoretical insights with algorithmic innovation, we contribute to both the understanding and the mitigation of risks in modern financial systems.

\section*{Materials and Methods}

\subsection*{Financial contagion in BIS data}

The Bank for International Settlements was founded in 1930 to facilitate reparations imposed on Germany after World War I. Reparation tasks became obsolete, and BIS became a provider of statistics and analysis that could aid the cooperation between central banks \cite{marcussen}. Due to globalization during this past century, there are growing financial inter-dependencies between developed countries, which has given BIS a renewed role. The BIS is now heavily involved whenever a financial crisis occurs, as it looks to recommend solutions to preserve financial stability \cite{hughes}. The institution expanded to 63 owner countries over the last three decades; it has become a meeting ground for regulators to pursue the standardization of banking practices and demand transparent reporting between the world's major financial players \cite{foster}.

Since the Latin American debt crisis of 1982, one of the topics of interest in BIS has been the inter-dependencies within the \emph{global lending network}; this is the worldwide debt held among banks from reporting countries. This data set is publicly available and published by the BIS Committee on the Global Financial System (or CGFS). While its official purpose is to identify potential sources of stress in the global financial system, it can also be studied to track where a banking crisis in one country can lead to a cascade of defaults in other countries; see \cite{cerqueti}. Previous studies conducted by Chen et al.\ explored financial contagion in the BIS network by sampling only 18 banking systems from this dataset; see \cite{chen1}. We will extract graphs from the entire global set of 62 CGFS's quarterly data reports from February 2000 to June 2015, modeling them as adversarial networks. Our goal will be to identify the presence of any low-key leaders and to determine if they play a significant role in the network's susceptibility to financial contagion.

The CGFS publishes two sets of international banking statistics every quarter. The first is for \emph{locational statistics}, which is a set of financial statements with assets and liabilities from banks in each reporting country. This information can be significant in the study of a country's internal financial stability. However, as we are concerned with the external inter-dependencies between countries, we omit locational statistics from our analysis. The second is a set of \emph{consolidated statistics}, which sums up the accumulated debt from all banks within a country and reports this sum as a consolidated debt-by-country balance. For example, if there are Canadian banks that have lent money through their head office or their overseas branches to banks that reside in France, then BIS collects the consolidated sum of debt that all French banks owe to all Canadian banks as a single amount. From a financial standpoint, this provides a direct measure of the \emph{risk exposure} each country's banking system has in an external crisis. To continue with our previous example, if the French banking system defaults, all Canadian banks with debt in French banks would now be exposed to the risk of defaulting on their obligations, triggering a potential crisis in the Canadian banking system.

Three types of consolidated statistics were published in the statistical annex of each BIS Quarterly Review. For our study, we will focus on the \textit{Foreign claims on an immediate borrower basis by BIS reporting country}, which were reported during the 2000 to 2015 year period under Table~\ref{tab:t1}. We chose this dataset for its simplicity. The table shows how much contractual debt one country's banking system holds over another. This was done through their banks' head offices and all their branches and subsidiaries worldwide on a consolidated basis, subtracting any internal bank amounts. For example, if a bank in Canada has branches in multiple countries, any debt that those foreign branches or subsidiaries hold will all still be reported as debt held by the Canadian bank. This is because the head office entity in Canada would bear the losses in the case of a default. Any debt held by a Canadian branch located in France is considered internal within the Canadian bank's financial statements. These are not to be counted as part of what debt France would owe Canada. The other two sets of consolidated data have been omitted from our analysis, as they include different attributes for risk mitigation, such as guarantees and collateral, which are tools to prevent the impact of defaults; these would add a higher level of complexity to the interpretations of our model and can be explored in future research.
\begin{table}[h!]
\caption{\textbf{Global banking network extracted from the BIS Quarterly Review Statistical Annex, June 2002.}}\label{tab:t1}
\centering
\begin{tabular}{ |p{2cm}|p{2cm}|p{2cm}|p{2cm}|p{2cm}| } 
 \hline
 Country &Austria&Belgium&Canada&Denmark\\
 \hline
 Austria & - & 3,179 & 1,467 & 179\\
 Belgium & 152 & - & 2,080 & 1,291 \\
 Canada & 300 & 1,845 & - & 123\\
 Denmark & 349 & 3,194 & 733 & - \\
 France & 1,665 & 43,141 & 4,742 & 827\\
 USA & 3,355 & 54,947 & 186,122 &   3,364\\
 UK & 4,191 & 62,365 & 34,328 &   9,781\\
 \hline
\end{tabular}
\begin{flushleft}
\textbf{Table notes:} Nodes represent countries, and directed edges indicate cross-border lending relationships, where an edge from country \( u \) to country \( v \) signifies that \( u \) owes money to \( v \). The edge weights correspond to the amount of debt in millions of US dollars. The network structure illustrates the flow of financial obligations between nations, with central lender countries playing a dominant role in global banking inter-dependencies.
\end{flushleft}
\end{table}

\subsubsection*{Adversarial networks}

\emph{Adversarial networks} are those in which edges model relationships involving competition, dominance, or enmity. We focus here on \emph{competition networks}: if $u$ is in direct competition with $v$, then the direction of the edge $(u,v)$ represents a negative correlation, such as when $u$ owes money to $v$. In \cite{bonato6,bonato2}, the authors developed a hypothesis that served as a predictive tool to uncover alliances and leaders within dynamic competition networks, where directed edges are added over discrete time steps.

Consider a dynamic competition network $G$. For nodes $u$, $v$, $w$, we say that $w$ is a \emph{common out-neighbor} of $u$ and $v$ if $(u,w)$ and $(v,w)$ are two directed edges in $G$. Let $\mathrm{CON}(u,v)$ be the number of common out-neighbors of two distinct nodes $u$ and $v$. The \emph{Common Out-neighbor} (or \emph{CON}) score of a fixed node $u$ is defined as 
\begin{equation}
\mathrm{CON}(u)=\sum_{v\in V} \mathrm{CON}(u,v).
\end{equation}
For a set of nodes $S\subseteq V(G)$ with at least two nodes, we define: 
\begin{equation}
\mathrm{CON}(S)=\sum_{u,v\in S} \mathrm{CON}(u,v).
\end{equation}

The CON centrality measure has been presented through multiple applications to dynamic competition networks in \cite{bonato6,bonato2}. A high CON score for a fixed node $u$ indicates that it shares many of the same adversaries with other nodes; hence, $u$ will influence how the network evolves. A low CON score means that a node does not share the same adversaries as other nodes, and therefore, it will not be as significant to the network's evolution. The \emph{Dynamic Competition Hypothesis} (or DCH) provides a quantitative framework for the structure of dynamic competition networks; see \cite{bonato6}. 

To apply this hypothesis to a banking network, we must introduce what an alliance and leader are within a competition network. \emph{Alliances} are associations formed for mutual benefit, such as countries that pool resources to achieve a common goal. In \cite{bonato2}, the authors study social game shows, such as Survivor, where alliances are groups of players who work together to vote off players outside the alliance. \emph{Leaders} would be players with a high standing among their peers in the network, and outgoing edges from these leaders will influence edges (which are votes) created by other players. The DCH states that dynamic competition networks arising from an adversarial network satisfy the following properties: (1) Strong alliances have low edge density; (2) members of an alliance with high CON scores are more likely leaders; and (3) Leaders exhibit high closeness, high CON scores, low in-degree, and high out-degree.

The DCH was tested against winners of social game shows, influential actors in political conflicts, and the hierarchical position of biological species in the food chain in \cite{bonato2}. The authors corroborated that alliances corresponded to near independent sets, that CON scores accurately determine leaders of alliances, and that leaders are detected via their CON scores and closeness.

A contrasting centrality measure for adversarial networks is PageRank, first introduced in \cite{brin}. For additional background on PageRank, see \cite{bonato}. A high PageRank for a fixed node $u$ in a directed graph will likely correlate with high in-degree nodes. This is why, for an adversarial network, we will compute the PageRank of nodes on a \emph{reversed-edge} network. Therefore, a high PageRank in an adversarial network will correlate with a node that has a high out-degree.

A \emph{low-key leader} (or \emph{LKL}) in an adversarial network is a node whose CON score and PageRank (edge-reversed) are negatively correlated, with a higher CON score and relatively low PageRank. LKLs were first discussed in \cite{bonato3}. Hence, an LKL node would still be influential in the network but with less centrality than a traditional leader would. To be able to calculate a difference between the two values, we note that CON-scores are positive integers and PageRank scores are probabilities, so we re-scale both scores by using \emph{unity-based normalization}:
\begin{equation}
X_{\mathrm{norm}}=\frac{X_i-X_{\mathrm{min}}}{X_{\mathrm{max}}-X_{\mathrm{min}}}.
\end{equation}
This scaling measure will satisfy $X_{\mathrm{norm}} \in [0,1]$ and can create a ranked order of nodes according to either their CON-score or PageRank.

Let $G$ be a directed graph, for the set of nodes $v_i\in V(G)$, where $1 \leq i \leq n$, and the CON-score and PageRank of $v_i$ are denoted by $\mathrm{CON}_i$ and $\mathrm{PR}_i$, respectively, we define \emph{low-key leader strength} as
\begin{equation}
\epsilon_i = \mathrm{CON}_{i,\mathrm{norm}} - \mathrm{PR}_{i,\mathrm{norm}},
\end{equation}
where $\epsilon_i \in [-1, 1]$. A node $v_L$ is a low-key leader if it has the maximum value of $\epsilon_L$, and $\epsilon_L>c$, where $c$ is a parameter determined by the network.

Specific to selected sets of adversarial dominance, cryptocurrency, and global trade networks, the authors in \cite{bonato3} identified low-key leaders as those with the highest values of $\epsilon$ and $\epsilon>0.5$. Further, it was shown that low-key leaders are present in most of the studied adversarial networks.

\subsubsection*{Applications to BIS data}

We obtained 62 tables corresponding to BIS Quarterly Review reports from February 2000 to June 2015. See Table~\ref{tab:t1} for an example of a data extract that shows where each row corresponds to a \emph{debtor} country, whose banking system owes money collectively to foreign \emph{lender} country's banks. Each column corresponds to the reporting lender country that holds the debt and is owed this amount. We construct a set of graphs $G_i$, where $i={1,2,\ldots ,62}$, by building 62 adjacency matrices from the data sets. Each node $v\in V(G)$ corresponds to a country and each weighted directed edge $(u,v;k)\in E(G)$ corresponds to the amount of money $k$ that is owed by country $u$ to the country $v$. The full data set of adjacency matrices is available on the following GitHub page: https://github.com/AdamSzava/BISNetworkData.

The directed edges $(u,v;k)$ are inherent adversarial relationships, where each debt $k$ from $u$ to $v$ is an adversarial obligation from a debtor to a lender counterparty. Graph $G$ is also a dynamic competition network as it evolves. Each of the reporting lender countries competes against the other to take advantage of new investment opportunities by adding new edges with outlier countries. Each of the debtor countries will compete against each other to obtain additional foreign investment from new edges with the lender countries.

We will compute the weighted PageRank on a reversed-edge network of each $G_i$, as well as the weighted CON Score of each node, to calculate the low-key leader strength of all countries for each of the 62 adjacency matrices.

\subsubsection*{The AML Model and Rabobank data}\label{aml}

The money laundering model has not evolved much since the Bank Secrecy Act. Even though law enforcement and regulators look at different aspects like the origin of the money and the intent of its use, the industry still defined policies around the 1980's placement, layering, and integration model \cite{cassella}. This model describes a method for concealing cash behind legitimate business activities. To money launder, the owner of the illicit profit will try to infiltrate the legal banking network through these three steps \cite{schneider}:
\begin{enumerate}
    \item \emph{Placement}: After obtaining illegal profit, the first step is to find a way for the owner to get the cash into the financial system, perhaps through multiple deposits or by commingling it with the proceeds of a legitimate, cash-intensive business, such as a restaurant or a gas station.
    \item \emph{Layering}: Once the cash is inside the financial system, the second step is moving the money through a series of transactions to make the trail difficult to follow.
    \item \emph{Integration}: The final step involves using the money in a legitimate transaction to pay for goods and services, whether to pay for the lifestyle of the original owner or to keep the operation going.
\end{enumerate}

With regards to combating placement, banks are required to maintain \emph{Know Your Client} (or KYC) profiles on each of their customers to understand the source of their income and determine what the intended use of the bank account will be \cite{nojeim}. Layering is the hardest step to combat, as banks currently monitor transactions from accounts flagged as \emph{high-risk}, so most supervised AML algorithms rely on existing flagged accounts. High-risk accounts are those being owned or controlled by individuals or corporations that have been linked to any of the following: economic sanctions, political exposure, cash-intensive businesses, regulatory or criminal enforcement, or money laundering scandals \cite{schneider,white}. Another strategy to combat layering relies on threshold reporting. Currently, most countries have a defined amount for all banks to report a transaction to the regulators; in the US and Canada, this amount is any transaction over \$10,000 in their respective currencies; see \cite{siacca}.

The algorithm we will introduce focuses on the layering and integration steps of money laundering within a network of accounts without having prior knowledge of any illicit placement or flagged high-risk accounts. A subset of results of this section were originally reported in \cite{bonAML}. The intent is that the algorithm will be capable of searching across a large data set of millions of bank account transactions to flag a potentially suspicious set of transactions that have moved less than the \$10,000 reporting threshold across different accounts, only to have returned a similar amount of money to the original account owner. It is worth noting that this does not achieve confirmation of money laundering. We are identifying a set of potentially suspicious accounts that must be further investigated. An additional purpose of our algorithm is that it can be used to run an initial identification step, which can then be further combined with one of the existing AML machine-learning algorithms. The added benefit is that our algorithm will flag a set of initial suspicious accounts within the real data instead of artificially introducing them. A parallel but unrelated approach using network science to detect financial crime was used in \cite{italy}.

Rabobank originated from 1,200 local cooperative banks, where volunteers managed local banking administration. The cooperatives’ governance structure, non-commercial ideology, and volunteer-based governance make them unique to other private banks in the country. In 1898, the Coöperatieve Centrale Raiffeisenbank and the Coöperatieve Centrale Boerenleenbank were founded to provide central organization and supervision, and then in 1972, they merged to form Rabobank \cite {degraaf}. Saxena et al.\ analyzed a network comprised of 1.6 million nodes with edges corresponding to transactions between the bank's users from the year 2010 to 2020; see \cite{saxena}. 

We used an anonymized Rabobank banking transaction data set provided by \cite{saxena}. This non-public data set consists of bank accounts and transactions between them. Each data element $\mathbf{\vec{p}}(u,v;n,k,Y_{1},Y_{2})$ has four different scalar labels. 
\begin{enumerate}
    \item Account Pair: A node pair of two accounts can denote a directed edge $(u,v)$ with a start node $u$ and an end node $v$.
    \item Number of Transactions: A positive integer $n$, the number of transactions between the two accounts in each pair.
    \item Amount of Money Transferred: An edge weight $k$, corresponding to the total amount of money in undisclosed \$ currency, transferred from the start node to the end node from 2010 to 2020.
    \item Start Year $Y_{1}$: The year corresponding to the first transaction for each directed edge.
    \item End Year $Y_{2}$: The year corresponding to the last transaction for each directed edge.
\end{enumerate}

\begin{figure}[ht!]
\centering
\includegraphics[width= 0.8\textwidth]{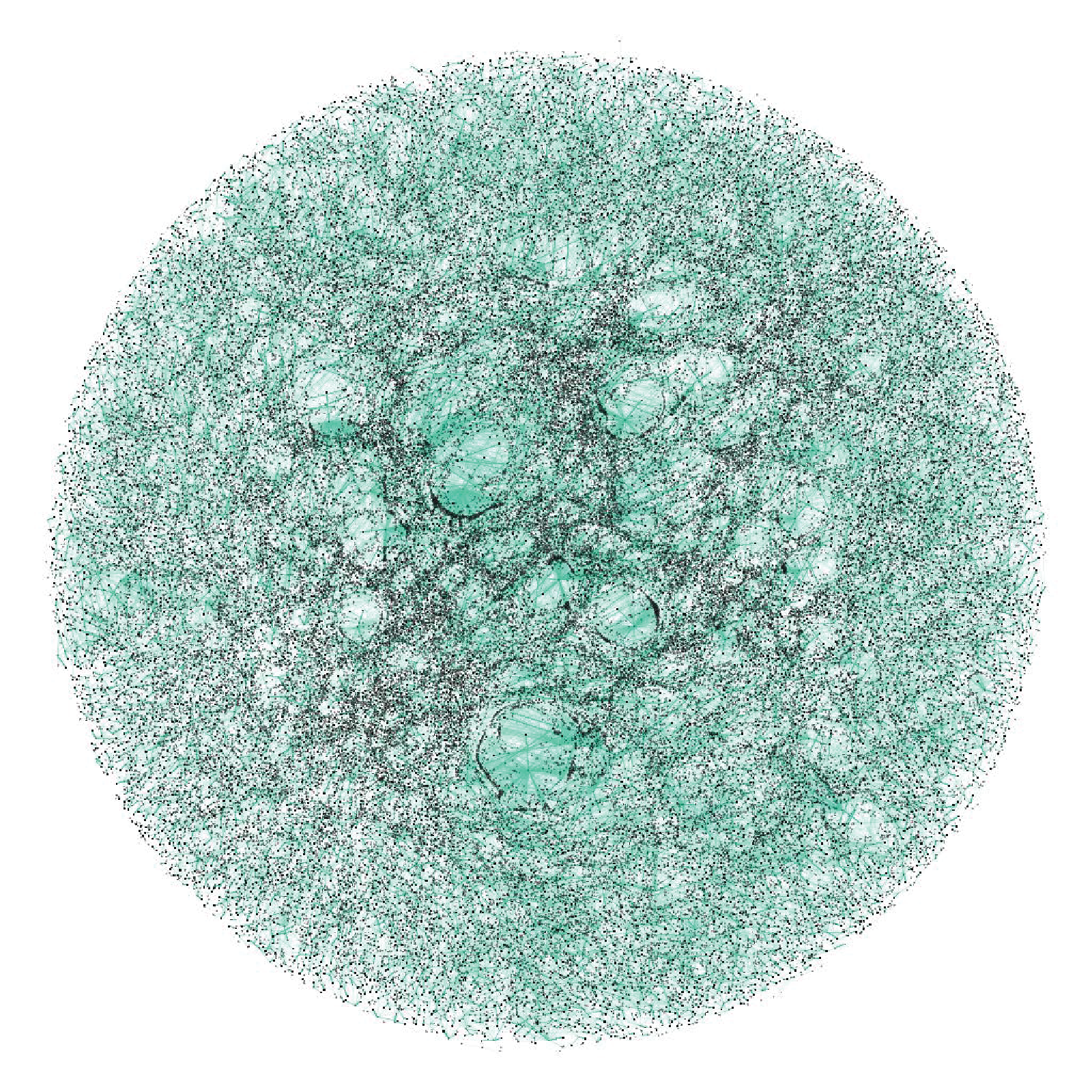}
\caption{\textbf{A visualization of a community from the Rabobank banking network, with transactions taken in the same year and average under \$10,000}.}\label{fig:fN}
\end{figure}

The data set comprises 1,624,030 nodes (which represent accounts) and 4,127,043 transactions. Since various transactions were done from the same start node to the same end node, there are only 3,823,167 edges. Saxena et al.\ studied the network’s community structure and showed that the bank's users are organized into smaller groups where some are making more transactions between them and fewer transactions outside the group; see \cite{saxena}. As discussed, a bank would report any transaction over \$10,000 due to regulatory thresholds. In Fig~\ref{fig:fN}, we see an extracted subset of 481,769 edges of unreported transactions; these edges represent transactions that occurred within the same year and were also under the \$10,000 threshold. Identifying cycles within this subset could still represent a computationally hard problem, and it would not tell us much about the association between a group of nodes. This is why, for us to search within the entire set of 1.6 million accounts, we need an initial algorithm that can partition the network into communities.

Community structures in networks refer to certain observable groupings characterized by nodes that are more densely connected within their respective communities and sparsely connected with nodes in other communities. Community detection involves finding these groups of nodes or communities by partitioning a network, which can help reveal underlying patterns and organization within the data. We used the well-known Louvain algorithm for community detection, which is based on modularity; see \cite{blondel}. 

Due to the size of a banking transaction network, we will also need a cycle-detecting algorithm for the final step of our algorithm. There are multiple available options. We have used the \emph{simple\_cycles} algorithm available on Python libraries; for more information on the background of this algorithm, please refer to Gupta et al.\ \cite{gupta}. We will now describe the combined steps that can be used to identify unusual movement of money within communities that could be flagged as potentially suspicious transactions.

We want to identify transactions within a community that could be attempting to layer and integrate illicit funds within the banking network. Additionally, we focus on the transactions that are not already being reported to the regulatory agencies by excluding those transactions that are over the \$10,000 threshold. Note that communities with orders strictly less than three can be excluded from the analysis, as bi-directional transactions are common in banking and are not enough to flag an attempt at layering illicit funds.

For a network of banking transactions $G$ with data elements of the form $\mathbf{\vec{p}}(u,v;n,k,t)$, where a pair of nodes $u,v\in V(G)$ represent two bank accounts, and $t$ represents the period during which the two accounts performed an $n$ number of transactions amounting to $k$ currency transferred.

\smallskip

We then define our algorithm with the following steps:
\begin{enumerate}
    \item Run the Louvain Algorithm on $G$ to detect a set $\mathbf{C}_{1}$ of communities larger than two nodes.
    \item Induce a set of subgraphs $\mathbf{C}_2$ from each community $C_{1,i}$ that contains pairs of nodes with edges that have a period $t$ less than an established threshold $t_{0}$.
    \item Induce a second set of subgraphs $\mathbf{C}_3$ from each community $C_{2,i}$ that contains pairs of nodes with edges that have an average transaction amount of $k<\$10,000$.
    \item Run the \emph{simple\_cycles} algorithm to generate a set of directed cycles $\mathbf{C}_4$.
\end{enumerate}

We set the one-year window \((t_0)\) to capture layering transactions that commonly occur within a short time (consistent with many AML detection frameworks). The result provides a list of similar nodes within community structures, such as $u$ in $C_{4,i}\in \mathbf{C}_4$ that are contributing to the movement of an unreported amount of money across all other nodes in the cycle $v$ in $C_{4,i}\in \mathbf{C}_4$, only to have an equivalent unreported amount of money returning to $u$ within a period less than $t_{0}$.

We also consider directed shortest paths within each subgraph in $\mathbf{C}_3$ that has at least one cycle. Directed paths are sequences of directed transactions that can be traced between two vertices but which (unlike in the case of directed cycles) do not return the funds to the source. Specifically, we are interested in \emph{long} shortest directed paths between nodes, which could also be an indication of an attempt at layering illicit funds. Once we identify the accounts that take part in these directed paths, we consider the correlation between these results and the results in \textit{Step 4}.
\begin{enumerate}
    \item[5.] For each subgraph in $\mathbf{C}_3$, if it contains at least one directed cycle, then add it to $\mathbf{C}_5$; otherwise, ignore it.
    \item[6.] For each subgraph in $\mathbf{C}_5$, compute the shortest directed distance between each pair of nodes using \textit{all\_pairs\_shortest\_path} in NetworkX. Generate a set of paths $\mathbf{P}_{\ell}$ with length $\ell$ such that $a \leq \ell \leq b$ for some choice of threshold values $a \text{ and }b$.
    \item[7.] For each subgraph in $\mathbf{C}_5$, let $P$ denote the set of nodes that are in at least one path (from \textit{Step 6}), and let $C$ denote the set of nodes that are in at least one directed cycle (from \textit{Step 4}). Compute the following $R$-value:
    \begin{equation}
    R = \frac{|P \cap C|}{|P|} = \frac{\text{\# of nodes in both a directed cycle and a directed path}}{\text{\# of nodes in a directed path}}.
    \end{equation}
\end{enumerate}

We focus on path lengths four to seven, inclusive, as these typically reflect more intricate layering schemes beyond trivial, short-hop transfers. Given a random node that is on a flagged directed path (that is, a directed path of length within the threshold values), the $R$-value represents the conditional probability that it is also on a directed cycle. A value closer to 1 would indicate a better correlation between these two directed structures.

\section*{Results}

\subsection*{BIS data}

Countries with a high-weighted PageRank correlate with nodes that have a high-weighted out-degree. These are countries that owe significant amounts of money to reporting lender countries. As a result of their high centrality, these countries are heavily integrated in the event of a financial contagion and become crucial to the stability of the network. If any of these countries default, then there is a greater risk to the entire network. During the period we studied, Germany, the United Kingdom, and the United States took their turns having the highest PageRank values. Some of the results for March 2006 are shown in Table~\ref{tab:t2}, where we can see that even though the United States owes the most amount of debt across the network, they do not have the highest PageRank during that same period.

\begin{table}[h!]
\caption{\textbf{Countries with the highest weighted PageRank in the BIS banking network for March 2006.}}\label{tab:t2}
\centering
\begin{tabular}{ |p{3 cm}|p{3 cm}|p{4 cm}| } \hline
 Country & $\mathrm{PageRank}_{\mathrm{norm}}$ & $\sum_{u}{|(v,u;k)|}$\\
 \hline
 Germany & 1 & \$1,077,765 \\
 United Kingdom & 0.9127 &  \$2,605,395  \\
 France & 0.6921 &  \$735,913 \\
 United States & 0.4833 &  \$4,457,212 \\
 Netherlands & 0.4729 &  \$621,576 \\
 \hline
\end{tabular}
\begin{flushleft}
\textbf{Table notes:} The PageRank values are computed on a reversed-edge network, where a higher PageRank indicates greater exposure as a major lender. The total debt column represents the sum of all outstanding loans issued by each country’s banking system, measured in millions of US dollars. While the United States has the highest total debt, Germany holds the highest PageRank, suggesting it occupies a structurally influential position in global financial inter-dependencies.
\end{flushleft}
\end{table}

A high PageRank also represents a higher likelihood that the random walk would visit the node at any given time. This has an indirect implication in our adversarial reversed-edge network. Whenever a country falls into default, it is more likely that it would have owed money to the country with the highest PageRank, or through transitivity, it could have owed money to a country that ultimately owes money to the country with the highest PageRank. Consequently, the country with the highest PageRank is at an increased exposure to any default risk from any random member of the network.

Countries with a weighted high CON Score correlate with sharing similar debt obligations with other nodes; these countries owe large amounts of money to selected lender countries that most other nodes in the network also owe money to. Their debt is both large and diversified, as they share similar debt obligations as their neighbor nodes; hence, the actions of these countries can influence how the network evolves. A low CON score could mean that a country does not owe a large amount of money and that any significant debt that this country owes might not be towards some of the same countries that its neighbors owe money to. As seen in Table~\ref{table:3}, which compares some results for March 2006, Germany has lent an amount over US \$3 trillion at this point, which is three times as much as the United States has lent US one trillion, and slightly above the United Kingdom who are owed US \$2.5 trillion. This means that the network as a whole owes more money to Germany than to the United Kingdom or the United States on a consolidated basis, and the countries with the highest CON Scores are the most likely to be the ones owing this money to Germany. A high CON Score gives a country the highest influence over causing a financial contagion if they fall into default. It is worth pointing out that the United States is consistently ranked with the highest weighted CON Score throughout the entire 2000 to 2015 period.

\begin{table}[h!]
\caption{\textbf{Countries with the highest weighted CON Score in the BIS banking network for March 2006.}}\label{table:3}
\centering
\begin{tabular}{ |p{3 cm}|p{3 cm}|p{4 cm}| } \hline
 \hline
 Country & $\mathrm{CON}_{\mathrm{norm}}$ & $\sum_{v}{|(v,u;k)|}$\\
 \hline
 United States & 1 & \$1,027,440\\
 United Kingdom & 0.5731 &  \$2,572,260\\
 Germany & 0.2294 &  \$3,151,383\\
 Italy & 0.1822 &  \$360,084\\
 France & 0.1810 &  \$1,754,414\\
 \hline
\end{tabular}
\begin{flushleft}
\textbf{Table notes:} The CON Score measures the extent to which a country shares common debt obligations with others, indicating its influence on financial contagion. A high CON Score suggests that a country’s financial stability is closely tied to that of multiple nations. The total debt column represents the sum of all external liabilities owed to foreign lenders, measured in millions of US dollars.
\end{flushleft}
\end{table}
\subsubsection*{Low-Key and Highly-Exposed Leaders in BIS}

For the calculation of the low-key leader strength, we explored maximum values of $\epsilon$ and determined that the absolute maxima $\epsilon=0.6036$ occurs for the United States on September 2005, as well as the relative minimum and maximum LKL strength values based on interquartile ranges oscillate from $\pm 0.02$ to $\pm 0.08$. Therefore, we determine that a low-key leader for the BIS network exists whenever there is an extremely positive LKL strength outlier such that $\epsilon_L>c$, and $c=0.1$.

\begin{figure}[ht!]
\centering
\includegraphics[width= 0.8\textwidth]{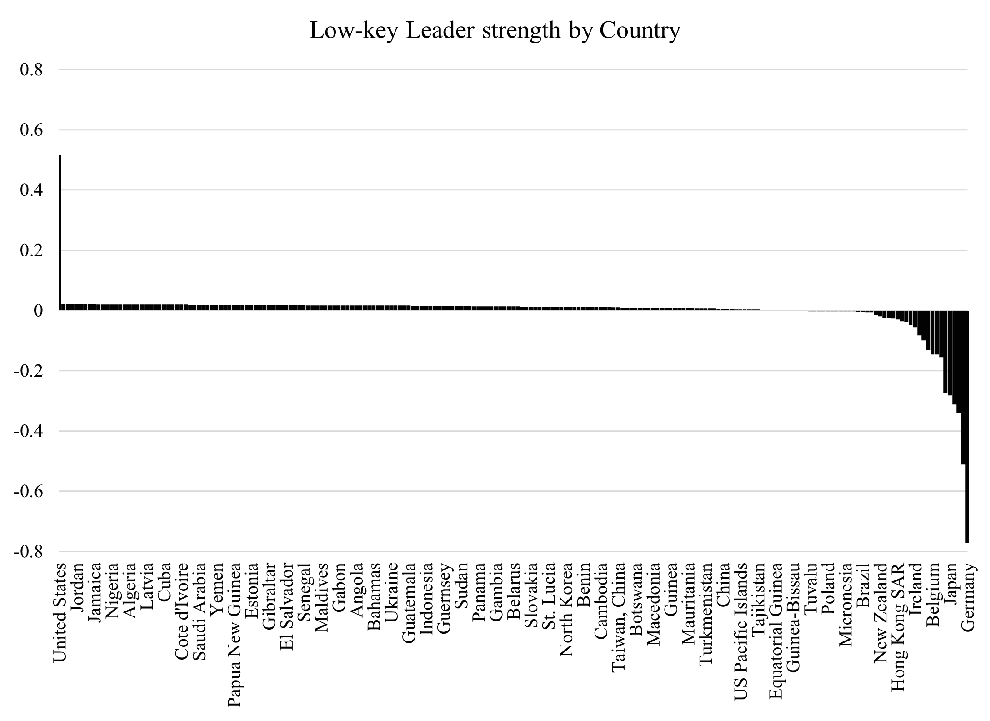}
\caption{\textbf{Low-Key Leader (or LKL) strength in the BIS network for March 2006.} \\
Countries with high positive LKL strength (such as the United States) maintain significant influence while mitigating exposure to financial contagion. In contrast, highly-exposed leaders, such as Germany and Japan, exhibit negative LKL strength, indicating heightened risk due to their extensive lending portfolios.}\label{fig:f1}
\end{figure}

During a brief period from February 2000 to December 2001, a few countries met this criteria, including the United States and the United Kingdom, notably Italy and Ireland. Throughout the period from March 2002 until December 2010, the United States was the only low-key leader that surpassed this threshold. From March 2011 through to June 2012, the United States and Mexico shared the status, until eventually, in September 2012, Mexico was left as the only low-key leader until June 2015. Fig~\ref{fig:f1} shows the low-key leader strength during March 2006, when the United States was the only low-key leader.

Being a low-key leader in the BIS network correlates with being one of the top lenders of money worldwide, while not necessarily the one at the very top with the greatest risk exposure. They could also have a smaller list of debtor countries than other top lenders. This country is one of the leaders that is sheltering itself from financial contagion by having a lower level of integration. A relatively higher weighted CON Score assures that they remain a top player with influence over the evolution of the network. The low-key leader owes large sums of money compared to other countries with similar PageRank, implying that they are the country that could cause the most damage if they were to default.

An interesting phenomenon also occurs at the other end of the LKL strength graph. Multiple countries are meeting the reversed threshold $\epsilon_L< -c$, see Germany and Japan in Fig~\ref{fig:f1}. This occurrence is not explored in the current literature, so we provide a possible interpretation of the significance of these nodes, which have a high-weighted PageRank and a relatively lower-weighted CON Score. In an adversarial network, define a \emph{highly-exposed leader} as one having a much higher PageRank than CON Score. Hence, a highly-exposed leader has negative low-key leader strength, typically below some fixed threshold $\epsilon < 0.$

We explored minimum values of $\epsilon$ and determined that the absolute minima $\mathrm{MIN}(\epsilon)=-0.7773$ occurs for Germany in June 2002, and while there's a consistent number of countries with $\epsilon$ values below -0.2, only a selected few have values below -0.4 across the entire period. Therefore, we determine that a highly-exposed leader for the BIS network exists whenever there is an extremely negative LKL strength outlier such that $\epsilon_L<C$, and $C=-0.4$. Germany and France are highly-exposed leaders consistently throughout the timeline. Japan briefly joined during the early 2000's and again after 2013, while the United Kingdom, which initially had high LKL strength, joined the highly-exposed leader group permanently after 2003.

Contrary to the low-key leader, having a negative LKL strength implies that this country has either lent money to the largest list of debtor countries or has the largest amount of money owed to them. This country is, therefore, heavily exposed to the risk of a financial contagion in the network. The relatively lower weighted CON Score also means they owe less money than other countries with similar PageRank. Owing less money to some of the major lender countries implies this country is less influential in the evolution of the network.

We next provide interpretative results for selected countries with high positive low-key leader strengths across the timeline studied, as well as some countries with extremely negative LKL strength values corresponding to highly-exposed leaders. We have also compared our measures of network centrality to a ranking based on a traditional macroeconomic indicator, the expenditure-based nominal \emph{gross domestic product} (or GDP): \begin{equation}
GDP=C+G+I+NX,
\end{equation}
where $C$ is consumption, $G$ is government spending, $I$ is investment, and $NX$ stands for net exports \cite{curtis}. The nominal GDP is a measurement of economic output that we can use to rank the list of countries in the BIS network by their productivity. To perform a direct comparison, we performed the same unit normalization on GDP. The result is an evaluation of a country's relevance to the network based on the measures of centrality: LKL strength, CON Score, PageRank, and nominal GDP, which provides insight into the actual size of their economies. 

\subsubsection*{United States}
\begin{figure}[ht!]
\centering
\includegraphics[width= 0.8\textwidth]{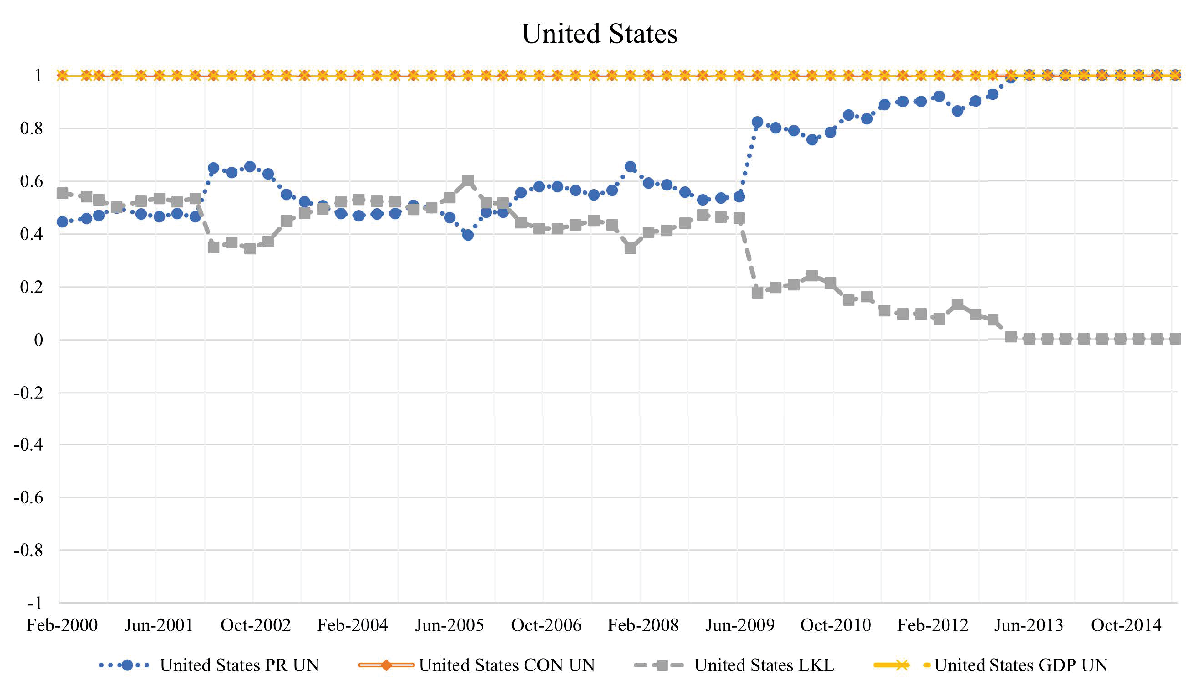}
\caption{\textbf{Evolution of PageRank, CON Score, LKL strength, and GDP for the United States in the BIS network.}\\ 
The PageRank (computed on the reversed-edge network) reflects the extent to which the U.S.\ banking system is a major lender, while the CON Score indicates its shared debt obligations with other countries. The LKL strength highlights periods when the United States exerted significant influence with relatively lower direct exposure to contagion risk. GDP is included for comparison, demonstrating how financial integration and systemic risk evolved alongside economic performance.}
\label{fig:fUSA}
\end{figure}

Most of the major economies are heavily invested in U.S.\ treasury bonds, as they are considered highly stable \cite{arvind} by having the highest weighted CON Score, the United States has the largest debt in common with its neighbors; this implies that if they were to default first on their debt obligations, they would cause the largest financial contagion to the lending network. They oscillate between other high PageRank countries up until the end of 2008, as seen in Fig~\ref{fig:fUSA}, until around 2009, which coincides with the global financial crisis, they begin to raise their integration to the network, until overtaking the highest PageRank position by the end of 2012. The United States was our low-key leader from 2000 to 2012 as they were not the most exposed to external risk during that period. Yet, they were always, you could say unfairly, the country that holds the most influence, as it is exposing the entire network the most to risk if they were to default. While the United States has historically been a key financial hub, studies suggest that it is not immune to cross-border contagion. The 2008 financial crisis demonstrated how financial instability in the U.S.\ banking sector had widespread global repercussions.

Their LKL strength began to drop after the 2008 global financial crisis so that we can consider a few possible explanations. Following the 2008 financial crisis, investor confidence in the U.S.\ economy remained strong, as evidenced by the continued high demand for U.S.\ Treasury securities. While financial markets faced uncertainty, U.S.\ Treasuries were still viewed as a global haven; see \cite{nippani}. However, the 52-week U.S.\ Treasury bond yield dropped below 1.0 percent during the financial crisis and remained low for an extended period, though it fluctuated and began to recover after 2013; see \cite{trading}. If we had expected that fewer countries would want to invest in the U.S., we would have seen its weighted CON Score decrease. However, by the end of 2012, they had the highest weighted PageRank and the highest weighted CON Score. We then consider a hypothesis of re-balancing, since after 2012, the United States is as equally exposed to foreign default risk as it is exposing the network to its own default risk. This re-balancing would align with the fact that they consistently remained the largest economy by GDP throughout the entire period, as seen in Fig~\ref{fig:fUSA}. 

We look for data that can hint towards a large increase in foreign investment; however, the U.S.\ Bureau of Economic Analysis shows no abrupt increase in direct investment abroad during 2012. To the contrary, it remains steady throughout the whole period; see \cite{usbureau}. Therefore, if the drop in LKL strength is not due to a change in the weighted CON Score, then it is solely due to an increase in weighted PageRank. This implies the U.S.\ has overall increased its exposure to the whole network after 2009, and this would only happen if the re-balancing occurred externally through the overall diversification of risk in the network. This last hypothesis does align with the following observation: the average LKL strength of all the countries with positive LKL values (besides the US) was 0.016 at the end of 2006, and it later increased to 0.021 by the end of 2014, as more nodes diversify to whom they lend money. This is also supported by studies that suggest portfolio re-balancing in European markets was a consequence, and a contagion transmitter, of the 2008 global financial crisis; see \cite{horta}.

\subsubsection*{Mexico}
In the lead-up to 1982, Brazil and Mexico were two major recipients of large volumes of foreign investment \cite{alvarez}, and the domestic banking sector of debtor countries like Mexico was involved in borrowing from international banking markets. While the BIS monitored international banking activities during the 1980's, the granularity and scope of data collection were not as comprehensive as in later years, potentially limiting detailed insights into the debt volumes and exposures of countries like Mexico during that period. When the Mexican system collapsed in August 1982, known as the Latin American financial crisis of 1982, it triggered a drive for improving policies and reporting for investments in developing countries due to the evidenced impact on international credit market stability; see \cite{howell}.

\begin{figure}[ht!]
\centering
\includegraphics[width= 0.8\textwidth]{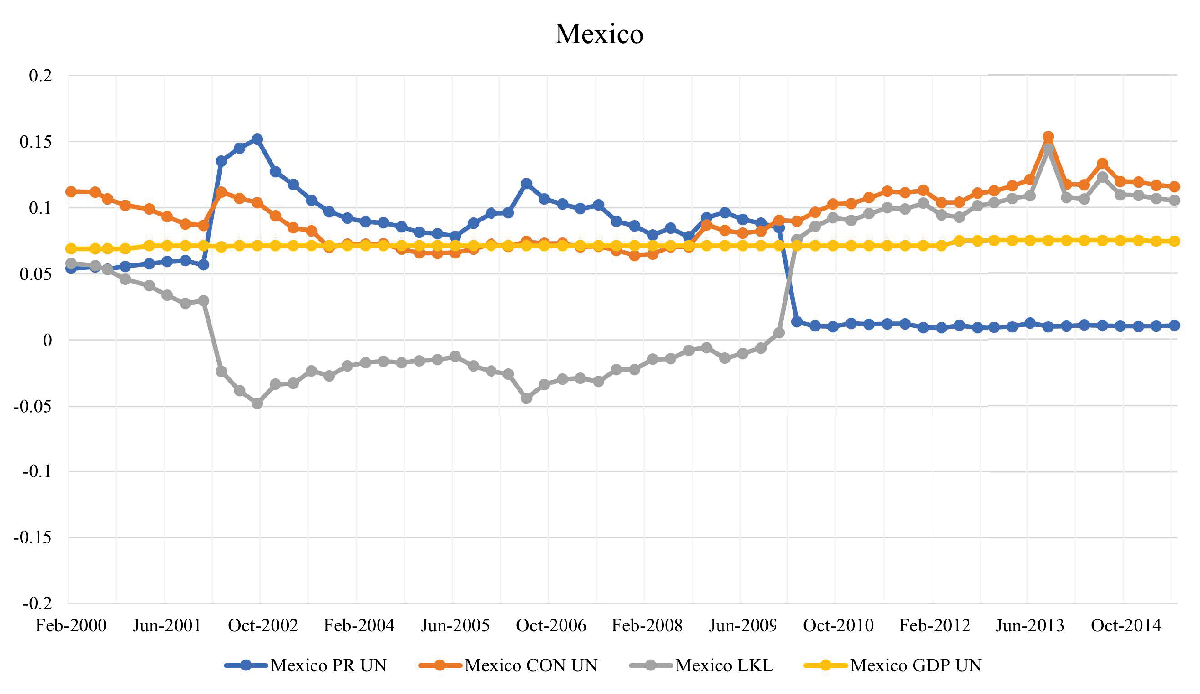}
\caption{\textbf{Evolution of PageRank, CON Score, LKL strength, and GDP for Mexico in the BIS network.}}\label{fig:f3}
\end{figure}

Refer to the values shown for the period 2000-2002 in Fig~\ref{fig:f3}, Mexico has a higher $\mathrm{PageRank}_{\mathrm{norm}}\approx0.1$ than $\mathrm{CON}_{\mathrm{norm}}\approx0.05$, meaning LKL strength is relatively high in the year 2000 but not enough for Mexico to be deemed a low-key leader. There is a large increase in weighted PageRank during the year 2001-2003, which coincides with the recession experienced by the Mexican economy in 2001-2003 \cite{erquizio}, and is claimed to have originated in the U.S.\ and propagated to the rest of the world \cite{reyes}. During this time, Mexico's weighted CON Score also increased as its total debt in the BIS network went from \$US 62 billion in December 2001 to more than \$US 211 billion in September 2002.

The Mexican economy entered a second recession during the 2008-2009 period \cite{erquizio}, and shortly after, the weighted PageRank rapidly declined during March 2010. In 2002, the BIS established a representative office in Mexico City, enhancing the collaboration between the BIS and the Bank of Mexico, which facilitated more detailed reporting on international banking activities, including data on debts owed to Mexican banks by other countries' banking systems. The weighted CON Score does not change as the debt they owe to other lender countries remains steady from \$US 494 billion in December 2009 to \$US 473 billion in March 2010. The increase in LKL strength may be associated with Mexico's financial interactions with various international financial centers, including jurisdictions like the Cayman Islands, which are known for their significant roles in global finance. Mexico has diminished its risk exposure from external defaults, and the decrease in PageRank is caused by lending to countries with lower integration in the network. The data shows that other countries, like the United States and Spain, remained heavily invested in Mexico after 2010, and they were exposed to a Mexican banking failure because a higher weighted CON Score could potentially cause financial contagion. Throughout the final period of 2012-2015, Mexico was the low-key leader of our network by surpassing the 0.1 threshold in September 2012, implying that countries in the BIS network are heavily exposed to a bank default in Mexico, while the sustained lower PageRank implies that Mexico has relatively shielded itself from external default risk.

\subsubsection*{United Kingdom}
\begin{figure}[ht!]
\centering
\includegraphics[width= 0.8\textwidth]{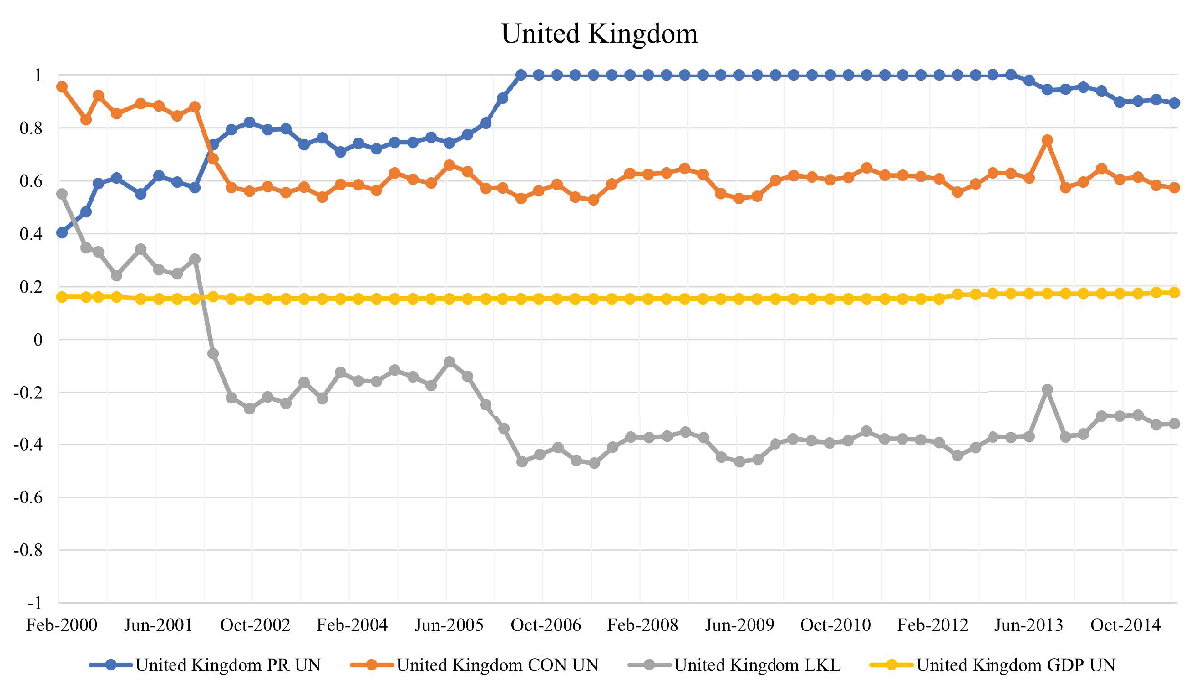}
\caption{\textbf{Evolution of PageRank, CON Score, LKL strength, and GDP for the United Kingdom in the BIS network.}}\label{fig:f4}
\end{figure}

The United Kingdom is a notable example of a country that transitioned from a low-key leader in 2000 to a highly-exposed leader by mid-2006. As seen in Fig~\ref{fig:f4}, this is due to the large difference between $|\mathrm{CON}_{\mathrm{norm}}|\approx0.85$ and $|\mathrm{PageRank}_{\mathrm{norm}}|\approx0.5$; which yields an average low-key leader strength of +0.35 during the first couple of years from 2000 to 2002. The UK has the second highest weighted CON Score throughout the entire timeline, which aligns with some of the conclusions from simulations in \cite{chen2}, which show that the United Kingdom, along with the United States, are the most important nodes that can cause a financial contagion in the BIS network.

The weighted CON Score slightly decreased after 2002, and the UK became a highly-exposed leader with an LKL strength of -0.45 by 2006. BIS Data shows that the amount of debt owed to other countries by UK banks increased slightly from \$US 971 billion in December 2001 to \$US 1,261 billion in September 2002, while the amount of money that the UK banking system owed doubled from \$US 536 billion to \$US 1,175 billion respectively. This growth in exposure during the 2005-2014 period can be correlated with the foreign expansion of the UK banking system; see \cite{faia}. In 2006, HSBC was among the largest banks in the world based on net assets \cite{riley}, which is the year the UK overtook the US and has the highest weighted PageRank in the BIS network. The global financial crisis would slow down this growth, along with attempts at tightening regulations by the Bank of England and the Financial Conduct Authority after 2008; see \cite{bell}.

\subsubsection*{Italy and Ireland}
\begin{figure}[ht!]
\centering
\includegraphics[width= 0.8\textwidth]{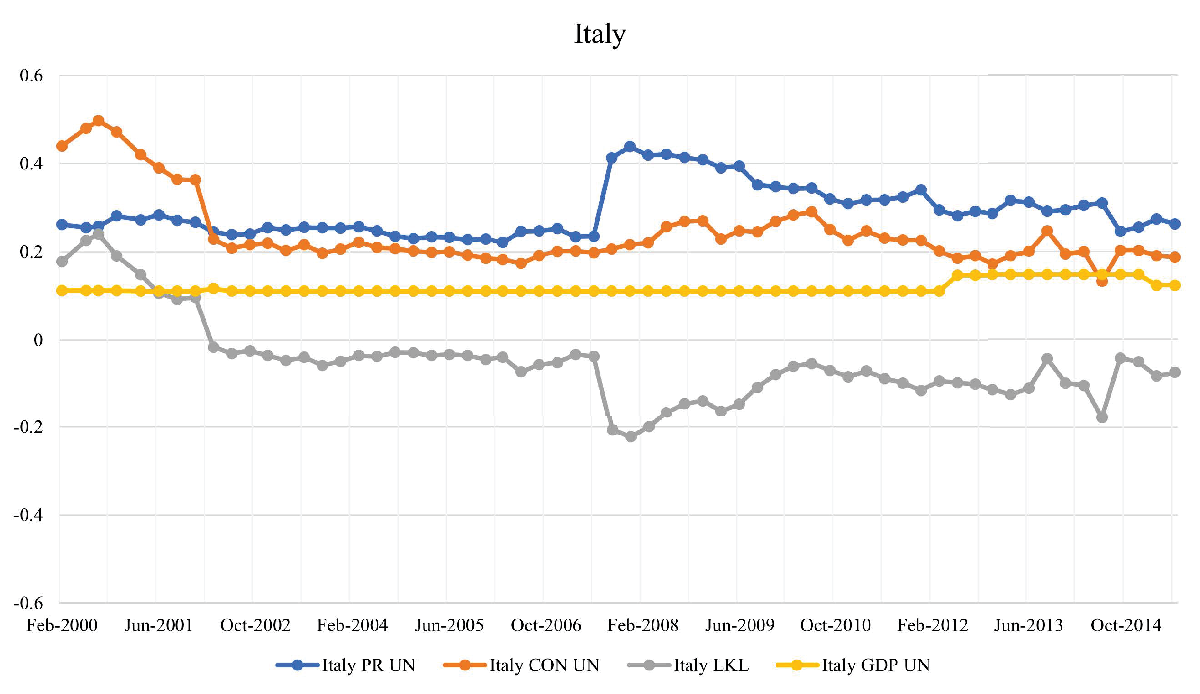}
\caption{\textbf{Evolution of PageRank, CON Score, LKL strength, and GDP for Italy in the BIS network.}}\label{fig:f5}
\end{figure}

\begin{figure}[ht!]
\centering
\includegraphics[width= 0.8\textwidth]{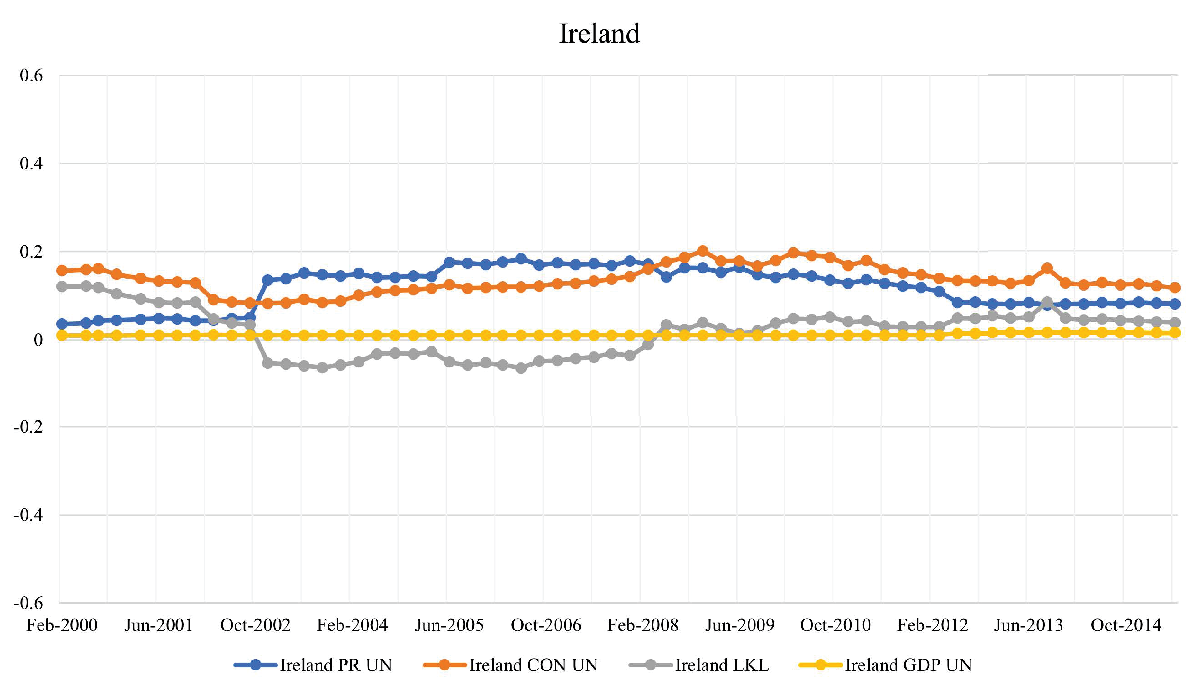}
\caption{\textbf{Evolution of PageRank, CON Score, LKL strength, and GDP for Ireland in the BIS network.}}\label{fig:f6}
\end{figure}

Italy and Ireland started as low-key leaders during the first two years of the studied timeline; see Fig~\ref{fig:f5} and Fig~\ref{fig:f6}, respectively. By the end of 2001, they had trailed back to lower than the $0.1$ LKL strength threshold. Italy maintained a higher exposure within the financial network leading up to the 2008 financial crisis. This exposure was due to a combination of structural economic challenges and high public debt levels. The collapse of Lehman Brothers in September 2008 further exacerbated these vulnerabilities, leading to a tightening of credit conditions by foreign lenders operating in Italy; see \cite{albertazzi}. The post-collapse decrease of credit extended by foreign banks in Italy could show why a wider gap occurs between PageRank and CON Score, aligning Italy towards the highly-exposed leaders.

Ireland's $\mathrm{PageRank}_{\mathrm{norm}}$ and $\mathrm{CON}_{\mathrm{norm}}$ rankings remained below 0.2, with a minimal difference between them, indicating that Ireland was neither a low-key leader nor a highly-exposed leader during the studied period. However, studies have shown that Ireland's economy is significantly integrated into the global financial system, primarily due to substantial foreign direct investment from multinational corporations; see \cite{mccauley}. This integration suggests that Ireland's exposure to external financial contagion may be more pronounced than some common measures of risk exposure indicate.

\subsubsection*{Germany}
\begin{figure}[ht!]
\centering
\includegraphics[width= 0.8\textwidth]{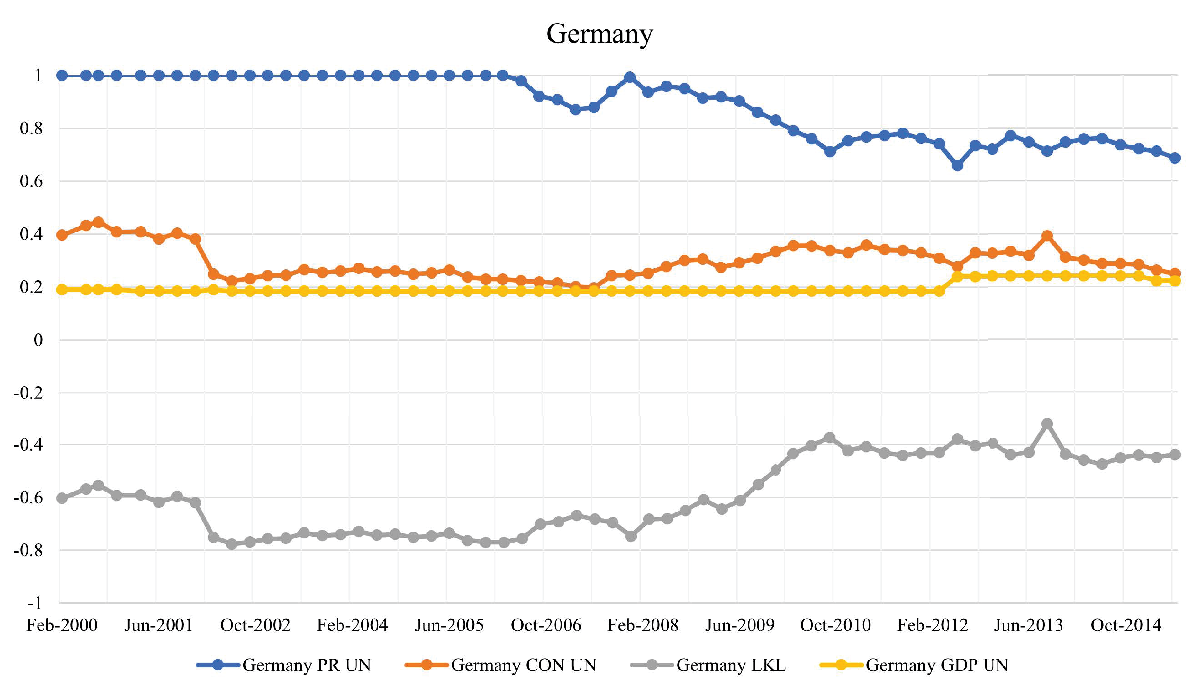}
\caption{\textbf{Evolution of PageRank, CON Score, LKL strength, and GDP for Germany in the BIS network.}}\label{fig:f7}
\end{figure}

Germany has the lowest LKL strength with values below -0.7 throughout the 2000-2010 period, which makes it the top highly-exposed leader node in the network before the 2008 financial crisis. Germany's GDP declined by 4.7\% in 2009, reflecting the severe impact of the global financial crisis on its economy; see \cite{rinne}. It can be seen that Germany has the highest weighted PageRank in Fig~\ref{fig:f7}, which correlates with lending the most amount of money. The characteristic of a highly-exposed leader comprises a relatively lower weighted CON Score, implying Germany does not put the network at the greatest risk if it defaults. This is shown in a study conducted in 2022 by Nikkinen et al., as most smaller economies across Europe were directly affected by the US and not by Germany; only Slovakia's economy was directly affected by Germany at a 5\% level. See \cite{nikkinen}.

After the financial crisis, the value of LKL strength slightly increased to -0.4; while Germany remains a highly-exposed leader, there is a significant decrease in PageRank, which is partially a consequence of the previously discussed overall diversification in the network, which ends up balancing the United States' position to take highest weighted PageRank by 2012. There is also a significant decrease in risk exposure to external defaults. At the beginning of 2007, Germany's banking system was owed \$US 4.1 trillion in debt worldwide; later, by the end of 2012, this number was brought down to \$US 2.7 trillion. The retreat is an overall sentiment felt across Germany's investors as the country tightened its regulations. In response to market conditions, certain German publicly-listed companies adjusted their strategies, with some opting to focus more on domestic investors and considering changes in their market listings; see \cite{bessler}.

\subsection*{AML and Rabobank data}

The AML algorithm described in the Methods section was implemented using Python on the Rabobank data set. Initial refinement activities mostly involved formatting the data to fit the requirements of each step in the algorithm. As the data set is not public, we will discuss the results of the different stages of the algorithm below.

\begin{enumerate}
    \item The initial data set was made up of 1,624,030 nodes and 3,823,167 edges.
    \item The Louvain algorithm returned 11,827 communities of account pairs and their corresponding edges. The average size of a community was 40, and the largest community contained 5,577 accounts.
    \item Out of the total number of edges within those communities, 1,763,113 edges corresponded to transactions conducted within one year or less.
    \item Out of that subset of transactions occurring within one year or less, 1,697,761 edges corresponded to transactions that amounted to an average transaction amount under the \$10,000 reporting threshold.
    \item After running the $simple\_cycles$ algorithm within those communities, we identified $83$ cycles of unreported transactions across $42$ communities, consisting of a total of $155$ nodes.
    \item After detecting directed paths (of length $4 \leq \ell \leq 7$) within communities that contained at least one cycle, we found $412,852$ paths, consisting of $7,527$ nodes, $146$ of which were in both a path and a cycle.
\end{enumerate}

The 83 generated cycles break down into the following: three 6-account cycles, two 5-account cycles, twenty-seven 4-account cycles, and fifty-one 3-account cycles. In total, 155 accounts are linked to unreported money transfers that initiate in one account and, after a certain number of transactions, return a similar amount to the initial account. 

\begin{figure}[ht!]
\centering
\includegraphics[width= 0.8\textwidth]{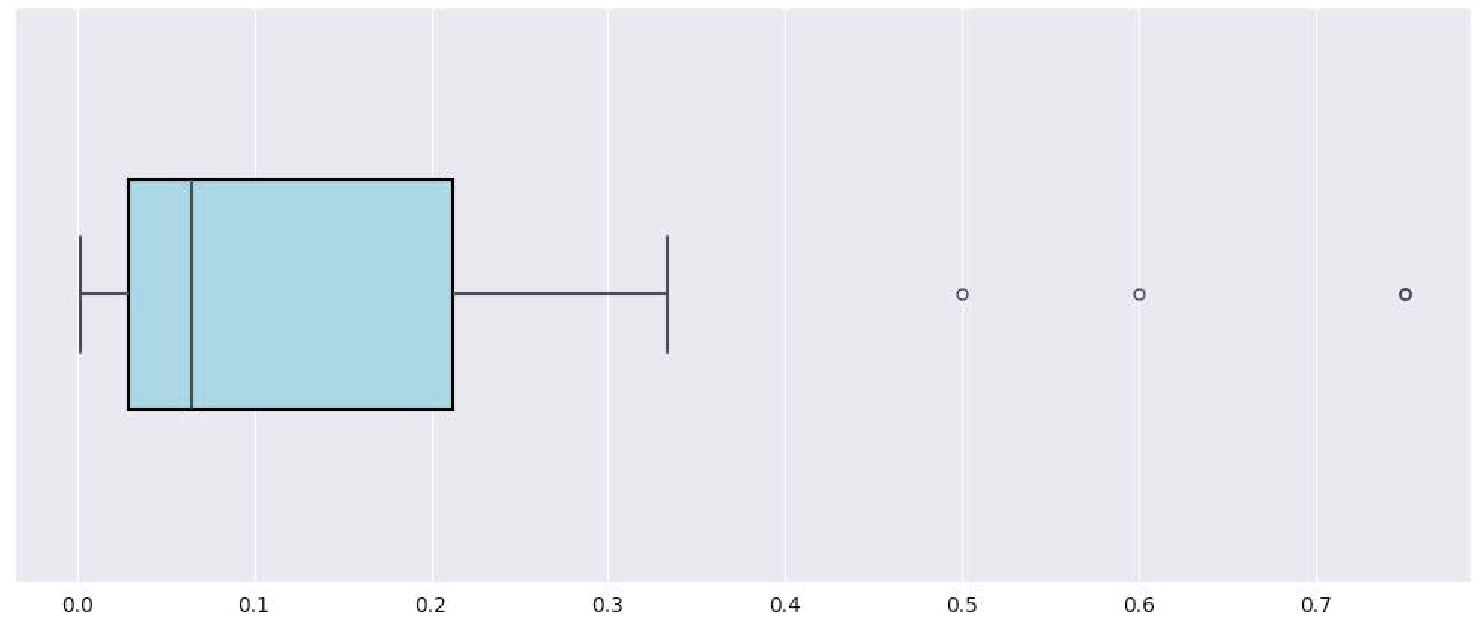}
\caption{\textbf{Distribution of R-values across the $42$ communities which contained at least one cycle.}}\label{fig:f9}
\end{figure}

Recall that the R-value in a community is the conditional probability that a randomly selected node that is on a directed path is also on a directed cycle. The results relating to the R-values are presented in Fig~\ref{fig:f9}. All but five communities scored a value less than $0.5$. Notably, three communities scored an R-value of around $0.75$.

This behavior can be flagged for further inspection, as performed through the existing suspicious transaction reports investigated by the bank's AML team to identify both the source of funds from these accounts or the purpose of the cyclic transactions. Even though 155 accounts may be viewed as a relatively large number to investigate, it is practical in the current banking industry, where thousands of transactions are analyzed yearly. Note that this could be a useful subset of accounts if determined to be flagged as high-risk. We can use this as an initial condition for one of the previously discussed machine learning programs. Furthermore, we could prioritize the three communities that had exceptionally high R-values, possibly indicating the involvement of more sophisticated money laundering operations. 

\section*{Discussion}

We demonstrate how network analysis can address two critical challenges in the global banking system: understanding systemic vulnerabilities and detecting potential financial crime. By analyzing BIS data, we identified low-key leaders and highly-exposed nodes in the global banking network, offering insights into how systemic risk propagates during financial crises. These findings emphasize the dual nature of influence within financial systems, where certain nodes, despite appearing less central, can exert an outsized impact on network stability. Concurrently, we applied our AML algorithm to anonymized Rabobank data, which underscores the potential of network-based methods to detect suspicious transactions below regulatory thresholds, offering a scalable way to strengthen existing compliance frameworks.

Our analysis of low-key leaders and highly-exposed nodes provides new perspectives on systemic risk within the BIS network, aligning with previous research that highlights the importance of diversification and integration in mitigating financial contagion. By quantifying these roles, we contribute to the broader understanding of how systemic risk evolves in complex financial systems. We define low-key leaders as countries with low diversification among top lenders. Such leaders subtly place the greatest risk exposure on their peers yet shield themselves from contagion via lower integration levels. By maintaining a relatively higher weighted CON Score, low-key leaders shape the network's evolution. The United States, Mexico, Ireland, and Italy illustrate this phenomenon: they do not provide the highest lending volumes, but, particularly in the case of the United States, they could trigger the largest financial contagion if their banking systems defaulted. Finally, the 2008 global financial crisis affected countries unevenly, and a shift toward greater diversification caused the United States to lose its low-key leader status.

We also introduce the concept of a highly-exposed leader, which is the opposite of a low-key leader and is characterized by a large negative LKL strength value. Such a country carries the greatest risk of contagion within the network. Its relatively low weighted CON Score implies it owes less money than countries with similar PageRank values, making it less influential in shaping the network's evolution. Germany and the United Kingdom emerged as prime examples of highly-exposed leaders. Germany maintained the highest risk exposure throughout our study period, whereas the UK transitioned into this position later, following a rapid expansion of its banking system until 2008–2010. Although regulatory bodies in both countries took measures to mitigate risk in the aftermath of the global financial crisis, they still exhibit lower weighted CON Scores and relatively higher exposures compared to other leading economies. Further research into the foreign investment strategies of low-key and highly-exposed leaders may illuminate the policies that enable economies to shield themselves from global financial contagion, particularly when large shifts occur in PageRank or CON Score rankings.

Our AML algorithm presents a novel way to detect suspicious activity in large transaction networks. Unlike traditional methods that depend on fixed thresholds or supervised learning, our network-based approach uses community detection and cycle identification to expose hidden patterns of suspicious behavior. In the Rabobank dataset, this algorithm successfully pinpointed cycles and longer transaction paths, both potential indicators of money laundering. These findings underscore the utility of unsupervised network strategies in AML systems, particularly where labeled data are limited or unavailable.
In a practical compliance workflow, the subset of accounts flagged by our network-based approach must be integrated with additional KYC data, transaction metadata, and risk indicators; this would help to reduce false positives and prioritize those cases warranting deeper investigation.

While our study offers important findings, it also suggests multiple avenues for future research. First, the AML algorithm could be expanded to inter-bank data from global payment systems such as SWIFT. Incorporating data from multiple institutions may reveal cross-border patterns of suspicious activity and regulatory discrepancies. Second, long directed paths within transaction networks, which may obscure the origin of illicit funds, deserve further investigation. These paths could complement cycle detection, providing a more comprehensive toolset for AML efforts. Finally, a deeper analysis of the BIS data could clarify how the global banking network evolves. By examining changes in centrality, alliances, and leader roles, we can gain a more nuanced view of how financial systems respond to crises and regulatory measures. To further enhance interpretability for stakeholders, it is useful to link changes in PageRank or CON score directly to real-world policy decisions, thereby illustrating how shifts in these measures reflect evolving systemic risk or resilience at a country level.

Our findings underscore the power of network analysis for both systemic risk assessment and regulatory compliance. By blending theoretical insights with practical applications, we contribute to the expanding body of research at the intersection of network science and finance. Future studies in this field have the potential to bolster financial stability and counter financial crime, thereby reinforcing the resilience of global banking systems in an ever more interconnected world.

\section*{Acknowledgements}
The first author acknowledges support from an NSERC Discovery Grant.

\section*{Author contributions}
Conceived and designed the experiments: AB JP AS. Performed the experiments: JP AS. Analyzed the data:  AB JP AS. Contributed to the writing of the manuscript:  AB JP AS. Contributed theory:  AB JP AS.

\end{document}